\documentstyle[11pt,newpasp,twoside]{article}
\index{instructions}
\index{guidelines}

\def\edcomment#1{\iffalse\marginpar{\raggedright\sl#1\/}\else\relax\fi}
\reversemarginpar

\begin{document}
\title{Solid Bare Strange Quark Stars}
 \author{Renxin Xu}
\affil{School of Physics, Peking University, Beijing 100871,
China}

\begin{abstract}
The reason, we need three terms of {\em `strange', `bare'}, and
{\em `solid'} before quark stars, is presented concisely though
some fundamental issues are not certain. Observations favoring
these stars are introduced.
\end{abstract}

\noindent
{\bf 1. Hadronic stability: why strange?}

Strange (quark) stars are quark stars with strangeness. What's
strangeness? Some peculiar cosmic ray events were discovered in
1947, with a strange nature of ``being produced rapidly in pairs,
but decaying slowly independently''. These particles were thus
called ``{\em strange}'' particles, and a new quantum number,
``{\em strangeness}''($S$), was introduced.
In the standard model of particle physics, it is known that the
strangeness represents actually the existence of a new kind of
quarks, the {\em strange quark}, $s$.
In this model, a neutron is composed of three valence quarks (one
$u$p and two $d$own quarks), sea quark pairs ($u{\bar u},d{\bar
u},s{\bar s}$), and $g$luons, i.e., $n=\{udd,u{\bar u}d{\bar
d}s{\bar s},g\}$; a proton $p=\{uud,u{\bar u}d{\bar d}s{\bar
s},g\}$. A neutron (or proton) does not have strangeness because
of $S(s)=-1$ and $S({\bar s})=1$. Hadrons with three valence
quarks are called baryons (e.g., protons and neutrons).

And then, how to produce strangeness in a star if it is the
remains of an evolved main sequent star?
Atomic nuclei in normal stars are made only of the two nucleons
(proton and neutron), without strangeness. Two scenarios are
outlined for creating strangeness in dense stellar matter.
(1) The first one is in the hadronic degrees of freedom (i.e.,
hadrons as quasiparticles). Hyperons are baryons with strangeness,
in which one or more valence quarks are replaced by $s$-quarks
(e.g., $\Lambda=\{usd\}$, $\Omega=\{sss\}$). Nucleons with high
enough Fermi energy (e.g., in traditional neutron stars) may decay
into hyperons by weak interactions. Such nuclear matter with
strangeness is called as strange {\em hadronic} matter, and the
corresponding neutron stars are named as {\em hyperon} stars. The
study of hypernuclei may help us to understand the hyperon stars.
(2) The second one is in the quark degrees. {\em Suppose} that
quarks within hadrons are deconfined in high density (to form
quark-gluon plasma, or called quark matter), one may expect that
2-flavor quark matter (i.e., $u$ and $d$) appears in ``neutron''
stars. However, in case of $s$-quark mass being smaller than the
Fermi energy of $u$- and $d$-quarks, the system may become more
stable via weak-interaction, decaying into 3-flavor quark matter.
A further radical speculation is that bulk 3-flavor quark matter
is absolutely stable (the Bodmer-Witten's conjecture); strange
stars (e.g., Xu 2002a) are accordingly made of such strange {\em
quark} matter (SQM) with nearly equal numbers of the light quarks.
Note: an SQM core and outer nuclear matter may coexist over
macroscopic scale if bulk SQM is metastable ({\em mixed} stars).

And then, can hadrons (e.g., neutrons) in ``neutron'' stars be
deconfined?
The underlying theory of the interaction between quarks is
believed to be quantum chromodynamics (QCD), based on which it is
proved that the interaction is asymptotically free in the smaller
scale (another property of QCD could be color confinement in the
larger scale, which has not been proved yet).
Short distance on average between quarks is possible by two
methods: (a) creating virtual quark pairs in relativistic heavy
ion collisions, (b) crowding hadrons in compact stars. The former
is of temperature-dominated, and the later
baryon-density-dominated.
Therefore, possible quark-deconfinement in compact stars might be
a straightforward consequence of asymptotic freedom.

In summary, the exist of strange stars (SSs), a special kind of
quark stars, could be quite natural if bulk SQM is an absolutely
stable state of 3-flavor color interaction system, though heavy
quark (charm, top, and bottom quarks) degrees could be excited if
the chemical potential is much high (e.g., in an SS core where the
density is very high). It is still not certain that pulsar-like
stars observed are neutron or strange stars.
{\em No} solid evidence for pulsars being neutron stars (NSs) yet
(though many authors believe this conventionally)!

\vspace{2mm} \noindent
{\bf 2. Formation and evolution: why bare?}

Because of the significant mass differences between $s$- and $u$-
(or $d$-) quarks, electrons in SQM maintain global charge
neutrality. The electrons occupy a larger space ($\sim 10^3$ fm in
radius) than that of quarks since the electromagnetic interaction
is weaker than the color one. This results in a strong magnetic
field ($\sim 10^{17}$ V/cm) just above the quark surface. The
field could repulse (or support) positively charged nuclei, and a
crust with mass $\sim 10^{-5}M_\odot$ ($\sim 10^2$ m thickness)
may cover the SQM. No difference between the surfaces of crusted
SSs and NSs.

In case of existence of SSs, radio pulsars were not supposed to be
bare SSs (BSSs) until Xu \& Qiao (1998) addressed that BSSs can do
as radio pulsars, with 3 advantages.
Can the quark surface be exposed to the cosmos? If can, the quark
surface could be used to identify an SS (Xu 2002b).
(1) A crust can hardly survive the detonation flame during a
combustion of nuclear matter into quark matter, because of rapid
energy release and high temperature.
(2) Unless the spin period is $>10^3$ s and the accretion rate is
$>$ the Eddington one, an isolate SS could not be covered by a
crust through accretion during its lifetime.

\vspace{2mm} \noindent
{\bf 3. Their astrophysical appearances: why solid?}

Condensation in momentum space (e.g., 2SC, CFL, LOFF states) is
currently focused on in the study of quark matter with high
density but low temperature. However, can condensation in position
space occur?
No such a competition (i.e., condensation in {\em momentum} vs.
{\em position} spaces) happens in the electric superconductivity
due to the strong Coulomb repulsion between electrons.

Quantum effects dominate in an ideal gas if the thermal de Broglie
wavelength is larger than the mean distance between particles, but
the case is different if strong interactions participate.
The interaction may favor a condensation in position space (Xu
2003), which results in the formation of $n$-quark clusters in SQM
(Note: these clusters are {\em not} bag-like color singlet
hadrons; if so, the Bodmer-Witten's conjecture is violated).
The mean distance between $n$-quark clusters is $l\sim [n/(3n_{\rm
b})]^{1/3}$ for SQM with density $n_{\rm b}$. The distance $l\sim
1$ fm for $n=1$ and $n_{\rm b}=2n_0$, where $n_0=0.16$ baryons per
fm$^3$ is the nuclear saturation density.
The interaction between quark clusters could be of a well
potential, the depth of which should be $V_0>\sqrt{(\hbar
c/l)^2+m^2c^4}-mc^2$, with $m$ the mass of the clusters, in order
to describe classically the cluster gas.
Furthermore, if the thermal excitation is not enough,
$\sqrt{(\hbar c/l)^2+m^2c^4}-mc^2+kT\ll V_0$, with $T$ the
temperature, the clusters might be localized to form a solid
state.
In the non-relativistic (NR) approximation, this means
$\hbar^2/(2ml^2)+kT\ll V_0$.
Assuming the interaction is of Lennard-Jones potential,
$V(r)=-A/r^6+B/r^{12}$, we have $l=(2B/A)^{1/6}$ and
$V_0=-V(l)=A^2/(4B)$. Therefore the condition for a solid state of
SQM in the NR case is $A^2/(4B)-\hbar^2/(2m)(A/2B)^{1/3}-kT\gg 1$.

Recent experimental evidence for multi-quark ($n>3$) particles
(e.g, Bai et al. 2003), although still not being understood in QCD
(e.g., Jaffe \& Wilczek 2003), may increase the possibility of
quark clustering in SQM.
These multiquark hadrons may decay rapidly by strong interaction;
nonetheless, such clusters could be stable in SSs since their
decay into hadrons would be suppressed or forbidden if the
Bodmer-Witten's conjecture is correct.

Actually, there may be a few observational hints of such a solid
quark state.

{\em Thermal spectra without atomic features} --- Atomic lines are
not detected definitely yet in the thermal radiation (Xu 2002c).
It is natural to understand the thermal spectra observed in a
model of SSs with solid quark surfaces, where the thermal
radiation could be analogous to that of metals, for RXJ1856 (Zhang
et al. 2003) as well as for other sources (Zhu et al. 2003, in
preparation).

There are, in fact, some efforts to understand the thermal spectra
in the conventional model of NS atmospheres.
(1) The featureless spectra was suggested to be an indication of
the vacuum polarization effect (Ho \& Lai 3002), but a
satisfactory fit is not done in the model with the inclusion of
the polarization effect.
(2) A hypothetical plasma phase condensation transition was
suggested (Lai \& Salpeter 1997) for NS atmospheres with high $B$
or low $T$, and was applied to interpret the featureless thermal
spectrum of RXJ1856 (Turolla, Zane \& Drake 2003). However a
consistent study of the phase transition in thermodynamics is
still not done yet.
(3) Rapid rotating of an NS may smear a spectral line, but such an
NS may hardly become radio ``death'' (i.e., below death lines).
Additionally, a problem could be inherent in all of these efforts:
{\em How to calm down the magnetospheric activities} (e.g.,
AXP/SGR-like persistent and burst X-ray emission, radio emission
due to pair-plasma instabilities, etc.) for such an NS with strong
field or high spin frequency?

{\em Pulsar glitching and free-precession} --- The current model
for glitches involves neutron superfluid vertex pinning and the
consequent fluid dynamics. However, the pinning should be much
weaker than predicted by the model at lest for two radio pulsars
(PSR B1828-11 and PSR B1642-03), otherwise the vortex pinning will
damp out the precession on timescales being much smaller than
observed. In addition the picture, that an NS core containing
coexisting neutron vertices and proton flux tubes, is also
inconsistent with observations of freely precessing pulsars (Link
2003).
Theoretically, a definitive conclusion on the nature of vertex
pinning has not been reached yet due to various uncertainties in
the microscopic physics.
This discrepancy could be circumvented if radio pulsars are solid:
no damping in free-precession of solid stars, and glitches
reflecting the behaviors of global starquakes.
The stresses, which trigger a starquake, develop due to the
spindown or, possibly, to the frame-dragging effect.

The global starquake can also results in an exponential recovery
of postglitches. For a solid star, the angular frequency
$\Omega(t)$ and the moment of inertia $I(t)$, as functions of time
$t$, are governed by \{${{\rm d}(I\Omega)/{\rm d}t} = -\alpha,
\dot{I} = -\kappa (I-I_0), I_0 = I_0(\Omega)$\}, where $\alpha$ is
the external braking torque which is known for a star with certain
magnetic momentum, $I_0$ is the inertia moment of the star in
force-free equilibrium (i.e., no stress there), and the recovery
of the inertial moment $I$ is assumed to be at a rate being
proportional to $(I-I_0)$.
These equations are closed if the function $I_0(\Omega)$ is given,
which can be well approximated by calculating the Maclaurin
configurations since the density of an SS with mass $< 1.4M_\odot$
is almost uniform.
If we temporarily neglect $\alpha$, $I_0$ is a constant. One then
has $\delta I\equiv I-I_0=-\Delta I \exp[-\kappa t]$ from the
second equation, if a glitch occurs at $t=0$, with $\Delta I$ the
initial departure of inertia moment after the glitch (Note:
$I(0)<I_0$).
Therefore we have a glitch recover behavior of the form
$\Omega(t)-\Omega(0)\sim \exp[-\kappa t]$, the postglitch
relaxation observed.
It is worth noting that the superfluid vortex pinning and
unpinning could also work during a starquake if, besides
quark-clusters localized, superfluid free quarks exist too.

{\em Others} --- (1) Pulsars with submillisecond spin periods?
Rotating fluid stars are subject to $r$-mode instability, which
results in temperature-dependent minimum spin periods for SSs
(Madsen 2000), but a solid star can spin more fast. Solid SSs can
be identified if discovering pulsar spin frequency beyond the
$r$-mode critical one. (2) Starquake-induced magnetic reconnection
or the strange planet's collision could be responsible to the
bursting X-ray radiation of AXP/SGRs, while propeller accretion
results in their persistent X-ray emission.

\vspace{2mm} \noindent
{\bf 4. Conclusions}

The time for ``neutron'' star study hasn't been passed although
it's been over 70 years since the related idea appeared. Such a
kind of compact stars may not just be boring big ``nuclei'', but
could be composed by matter of a {\em new} state --- quark-gluon
plasma. Recent observations challenge the conventional NS models,
and should reveal valuable information of the quark matter state.

\acknowledgments This work is supported by NSFC (10273001) and the
Special Funds for Major State Basic Research Projects of China
(G2000077602).


\begin{references}

\reference Bai, J. Z., et al. 2003, \prl, 91, 022001

\reference Jaffe, R., Wilczek, F. 2003, preprint (hep-ph/0307341)

\reference Ho, W. C. G., Lai, D. 2003, \mnras, 338, 233
(astro-ph/0201380)

\reference Lai, D., Salpeter, E. E. 1997, \apj, 491, 270

\reference Link, B., 2003, \prl, 91, 101101 (astro-ph/0302441)

\reference Madsen, J. 2000, \prl, 85, 10 (astro-ph/9912418)

\reference Turolla, R., Zane, S., Drake, J. 2003, \apj, in press
(astro-ph/0308326)

\reference Xu, R. X., Qiao, G. J. 1998, Chin.Phys.Lett., 15, 934
(astro-ph/9811197)

\reference Xu, R. X., 2002a, Proceedings of IAU Symp. No. 214, in
press (astro-ph/0211348)

\reference Xu, R. X., 2002b, Proceedings of 6th pacific rim
conference (astro-ph/0211563)

\reference Xu, R. X. 2002c, \apjl, 570, L65 (astro-ph/0202365)

\reference Xu, R. X., 2003, \apjl, 596, L59 (astro-ph/0302165)

\reference Zhang, X. L., Xu, R. X., Zhang, S. N., this preceddings
(astro-ph/0310049)

\end{references}
\end{document}